\documentclass[useAMS,usenatbib]{mn2e}
\usepackage{graphicx,amssym}
\newcommand{\bc}{\begin{center}}
\newcommand{\ec}{\end{center}}
\usepackage{color}
\newcommand{\comment}[1]{}

\title[Starburst galaxies in SAMs]
      {Starburst galaxies in semi-analytic models of galaxy formation and evolution}

\author[Wang et al. ]
       { Lan Wang$^{1}$$\thanks{Email: wanglan@bao.ac.cn}$, Gabriella De Lucia$^2$,  Fabio Fontanot$^2$, Michaela Hirschmann$^3$\\      
        $^1$Key Laboratory for Computational Astrophysics, National
Astronomical Observatory, Chinese Academy of Sciences,\\
Datun Road 20A, Beijing 100012, China\\
	$^2$INAF - Astronomical Observatory of Trieste, via G.B. Tiepolo 11, I-34143 Trieste, Italy\\	
    $^3$Sorbonne Universit\`es, UPMC-CNRS, UMR7095, Institut d\' Astrophysique de Paris, F-75014 Paris, France
} 
\begin{document}

\pagerange{\pageref{firstpage}--\pageref{lastpage}} 
\pubyear{2018}

\maketitle

\label{firstpage}

\begin{abstract}
We study the shape and evolution of the star formation main sequence in three independently developed semi-analytic models of galaxy formation. We focus, in particular, on the characterization of the model galaxies that are significantly above the main sequence, and that can be identified with galaxies classified as `starburst' in recent observational work. We find that, in all three models considered, star formation triggered by merger events (both minor and major) contribute to only a very small fraction of the cosmic density of star formation. While mergers are associated to bursts of star formation in all models, galaxies that experienced recent merger events are not necessarily significantly above the main sequence. On the other hand, `starburst galaxies'  (defined as those with SFRs larger than 4 times the typical value in the main sequence) are not necessarily associated with merger episodes, especially at the low-mass end. Galaxies that experienced recent mergers can have relatively low levels of star formation when/if the merger is gas-poor, and galaxies with no recent merger can experience episodes of starbursts due to a combination of large amount of cold gas available from cooling/accretion events and/or small disk radii which increases the cold gas surface density. 
\end{abstract}



\section{Introduction}
\label{sec:intro}

Star forming galaxies exhibit a relatively tight correlation between star formation rate (SFR) and stellar mass, that is traditionally referred to as `main sequence'  \citep[MS;][]{noeske2007}.  This relation is observed in the local Universe \citep[e.g.][]{brinchmann2004,salim2007}, at z $\sim 1-2$ \citep[e.g.][]{noeske2007, rodighiero2011}, at z $ \sim 3-4$ \citep[e.g.][]{daddi2009, pannella2015}, and up to z $ \sim 7$ \citep[e.g.][]{drory2008, salmon2015}. 
 The correlation is tight, with a scatter of less than 0.3 dex \citep[e.g.][]{daddi2007,pannella2009, kurczynski2016}, 
which is typically interpreted in terms of smooth gas accretion and steady star formation \citep{noeske2007, renzini2009, finlator2011}.  Galaxies that reside significantly above the observed MS are usually classified as `starburst' galaxies and are believed to be driven above the MS by more violent processes like gas-rich mergers  \citep[e.g.][]{cibinel2018} or disk instabilities  \citep[e.g.][]{fathi2015, romeo2016}. Recent observations have revealed the presence of a significant starburst population up to $z\sim ~ 5$ \citep{rodighiero2011, schreiber2015, bisigello2017}, although recent studies by e.g. \citet{zhang2018} suggest that the SFR might be over-estimated in environments where the stellar initial mass function (IMF) is top-heavy.  

The relative importance of the starburst population varies in different studies, probably due to different selections and definitions adopted. \citet{rodighiero2011} and \citet{schreiber2015} find that up to  $z \sim 3$, starburst galaxies are rare, i.e they account for no more than 2-3 percent of the overall star formation population. The fraction of starburst galaxies remains almost constant up to redshift $\sim$3 for galaxies more massive than $5 \times 10^{10}M_{\odot}$ \citep{schreiber2015}. \citet{bisigello2017} claim that the fraction of starburst galaxies increases with decreasing stellar mass, and increases with increasing redshifts. In particular, for galaxies with $\log({\rm M}_{\rm stars}/{\rm M}_{\odot})=8.25-11.25$, the fraction of starburst galaxies raises from about 5 per cent at z = $0.5 -1.0$ to about 16 per cent at z = $2 - 3$. \citet{caputi2017} find that 15 per cent of galaxies more massive than $10^{9.2}M_{\odot}$ are starburst  at z = $4 - 5$.

Theoretical models of galaxy formation typically adopt empirical star formation prescriptions to describe the correlation between the amount of dense cold gas available and the star formation rate. Generally, star formation in disks is described following some variants of the observed Kennicutt-Schmidt law (this is the so-called `quiescent' mode for star formation), and events like mergers and disk instabilities are typically associated to starburst episodes (the `starburst' mode).  The detailed prescriptions adopted vary in different models  \citep[e.g.][]{white1991, somerville2001, lagos2011} and can be tuned using results from controlled numerical experiments \citep[e.g.][]{robertson2006, cox2008, somerville2008}. Both recent  hydrodynamical simulations \citep{finlator2006, dave2008, hirschmann2013, cochrane2018} and semi-analytic models \citep{lagos2011, mitchell2014, cowley2017,xie2017}  predict  a relatively tight SFR-stellar mass relation at different cosmic epochs. Most models, however, tend to under-predict the level of star formation observed at high redshift (but see recent discussion in \citet{fontanot2017}). 

No detailed analysis has been carried out yet on the relative contribution of the starburst population in these models.  One relevant exception is the recent work by \citet{wilkinson2018}, based on the Illustris Simulation, who find that about half of the `starburst' galaxies selected at present have not undergone a merger in the past 2 Gyr and argue that these are likely triggered by interactions with other galaxies at $z<1$.  In this study we use three semi-analytic models to quantify the relative importance of the starburst population at different redshifts, and their triggering mechanisms. 
Using semi-analytic models, we have the advantage to explore the effect of different physical processes on the starburst populations more efficiently than in hydrodynamical simulations. In addition, using independent models, we can give an assessment on the robustness of the results. 

The paper is organized as follows: Section 2 introduces the three models we use in this study. In Section 3, we present the basic statistics predicted, including the SFR function, the SFR density  evolution, the SFR - stellar mass relation, and the starburst fractions at different redshifts. In Section 4, we quantify the relative importance of recent mergers among the starburst galaxies, and in Section 5 we analyse the physical mechanisms leading to SFRs elevated with respect to the main sequence. Section 6 provides a discussion of our results and our conclusions.

\section{Galaxy formation models}
\label{sec:models}
In this section, we introduce the three semi-analytic models adopted in this study focusing on the prescriptions adopted to model star formation. More detailed and extensive descriptions of each model can be found in  \citet{DLB07}, \citet{faro2009} and \citet{hirschmann2016}, respectively (and references therein).

\subsection{DLB07}
This widely used model introduced in \citet[][hereafter DLB07]{DLB07} is based on the semi-analytic model developed by the Munich group \citep{kauffmann2000,springel2001,delucia2004,croton2006}.  The model has been run on subhalo merger trees extracted from the Millennium Simulation \citep{springel2005}. The adopted cosmological model is a flat $\Lambda$CDM cosmology with $\Omega_{\rm m}=0.25$, $\Omega_{\rm b}=0.045$, $h=0.73$,$\Omega_\Lambda=0.75$, $n=1$, and $\sigma_8=0.9$. The simulation follows $N= 2160^3$ particles of mass $8.6\times10^{8}\,h^{-1}{\rm M}_{\odot}$ from redshift $z=127$ to the present day, within a comoving box of $500\, h^{-1}$Mpc on a side.  Full particle data are stored at 64 output times. For each output, haloes are identified using a classical friends-of-friends (FOF) group-finder algorithm, while substructures (or subhaloes) within each FOF halo are found using the SUBFIND algorithm of \citet{springel2001}. Subhalo merger trees are constructed by defining a unique descendant subhalo at any given snapshot \citep{springel2005}. 

Star formation takes place in galactic disks, when the cold gas mass  exceeds a critical value $M_{\rm crit}$ \citep{kennicutt1998}, and the SFR is assumed to be proportional to the cold gas mass and inversely proportional to the disc dynamical time $\tau_{\rm dyn}$. Specifically:

\begin{displaymath}
 M_{\rm crit} = 0.19 \times 10^{10} (\frac{V_{\rm vir}}{\rm km \space s^{-1}}) ( \frac{r_{\rm disc}}{\rm Mpc} ) M_{\odot}, 
\end{displaymath}

and
\begin{displaymath}
{\rm SFR}_{\rm DLB07, disk} = 0.03 \times (M_{\rm cold} - M_{\rm crit})/\tau_{\rm dyn},
\end{displaymath}
where $\tau_{\rm dyn}$ = $r_{\rm disc}/V_{\rm vir}$, with $V_{\rm vir}$ representing the virial velocity of the host halo and $r_{\rm disc}$  the cold gas disc radius. The latter is approximated as three times the disk scale length \citep{croton2006}. 

When a galaxy merger occurs, a fraction $f_{\rm DLB07, burst}$ of the cold gas available is assumed to be converted into stars.  For major mergers (mass ratios larger than 0.3), this fraction is assumed to be equal to 1 while for minor mergers it is parametrized as \citep{somerville2001}:
\begin{displaymath}
f_{\rm DLB07, burst}=0.56\times(M_1/M_2)^{0.7}
\end{displaymath}

\subsection{\sc MORGANA}
The MOdel for the Rise of GAlaxies aNd Agns ({\sc MORGANA}) is described in detail in \citet{monaco2007} and \citet{faro2009}. {\sc MORGANA} has been run on FOF-based merger trees extracted from a {\sc PINOCCHIO}\citep{monaco2002} realization of a 200$^3$ Mpc$^3$ volume,  using $1000^3$ particles (particle mass $2.84 \times 10^8 M_{\odot}$) and assuming a Wilkinson Microwave Anisotropy Probe 3 \citep{spergel2005} cosmology with $\Omega_{\rm m}=0.24$, $\Omega_{\rm b}=0.0456$, $h=0.73$, $\Omega_\Lambda=0.76$, $n_s=0.96$, and $\sigma_8=0.8$. As discussed in previous comparison work, we do not expect the slight difference in cosmology with respect to the Millennium Simulation to affect significantly our results \citep{wangjie2008, guoqi2013}. 

In {\sc MORGANA}, a star formation timescale in disks is estimated from the cold gas surface density $\sigma_{\rm cold, disk}$ and the cold gas fraction $f_{\rm cold}$ in the disc component:
\begin{displaymath}
\tau_{\rm MORGANA,disk} = 9.1\times (\frac{\sigma_{\rm cold,disk}}{M_{\odot}{\rm pc^{-2}}})^{-0.73}(\frac{f_{\rm cold}}{0.1})^{0.45} {\rm Gyr}
\end{displaymath}
Unlike in DLB07, no critical cold gas threshold is assumed for stars to form, and star formation is also allowed in bulges (cold gas can flow into the bulge component as a result of galaxy mergers and disk instabilities). In the latter case, however, the time-scale for star formation is much shorter than that of the disk: 
\begin{displaymath}
\tau_{\rm MORGANA,bulge}=4 \times (\frac{\sigma_{\rm cold,bulge}}{M_{\odot}{\rm pc^{-2}}})^{-0.4} {\rm Gyr}
\end{displaymath}

In both cases, star formation is modeled assumed a Kennicutt--Schmidt relation: 
\begin{displaymath}
{\rm SFR}_{\rm MORGANA} = M_{\rm cold} /\tau_{\rm MORGANA}
\end{displaymath}
Since the gas in the bulge is more concentrated than that of the disk, it has relatively high gas surface densities and therefore relatively short star formation time-scales, effectively giving rise to `starbursts'.  Disk instability events are assumed to move half of the total mass (stars plus cold gas ) of the disk to the bulge component, providing an important channel for bulge formation\footnote{In DLB07 and GAEA, disk instabilities episodes are not associated to gas flows towards the center, and therefore to starbursts, as described in detail in e.g. \citet{delucia2011}.}. 

One important difference between {\sc MORGANA} and DLB07, that we identified in previous work \citep{delucia2010}, lies in the estimate of merger timescales.  When a galaxy becomes a satellite in MORGANA, it is assumed to merge with its central galaxy on a timescale computed using the fitting formulae provided by \citet{taffoni2003} that accounts for dynamical friction, mass loss by tidal stripping, tidal disruption of subhaloes and tidal shocks.  \citet{delucia2010} showed that this formulation results in merger times that are systematically lower than those assumed in DLB07, that implements a modification of the classical dynamical friction formula. Since mergers are a primary channel for driving starbursts, we will consider below  a realization of the MORGANA model using merger time-scales close to those assumed in DLB07.  We have checked that using the default merger timescales in standard MORGANA would result to in general higher fraction of merger galaxies, but would not change qualitatively the conclusions of our work. 

\subsection{GAEA}
The GAlaxy Evolution and Assembly (GAEA) model \citep{hirschmann2016} is a development of the DLB07, including (a) modifications to follow more accurately  processes on the scale of the Milky Way satellites as described in \citet{delucia2008}  and \citet{li2010}, (b) a sophisticated chemical enrichment scheme tracing the evolution of individual elements \citep{delucia2014}, and (c) a modified stellar feedback and gas recycling scheme partly based on results from hydrodynamical simulations \citep{hirschmann2016}.  

The modeling adopted for star formation is the same as in DLB07, with an update of the calculation of disk radii and of the critical surface density below which no star formation takes place (these are detailed in De Lucia \& Helmi 2008). The modified feedback and recycling scheme, however, implies significantly higher gas fractions at higher redshift with a modulation as a function of stellar mass that is in nice agreement with observed trends \citep{hirschmann2016}. It is therefore interesting to extend our analysis to this new model that has been applied to the same Millennium Simulation described above.

\section{SFR distributions as a function of cosmic times}

\begin{figure*}
\bc
\hspace{-1.4cm}
\resizebox{15.cm}{!}{\includegraphics{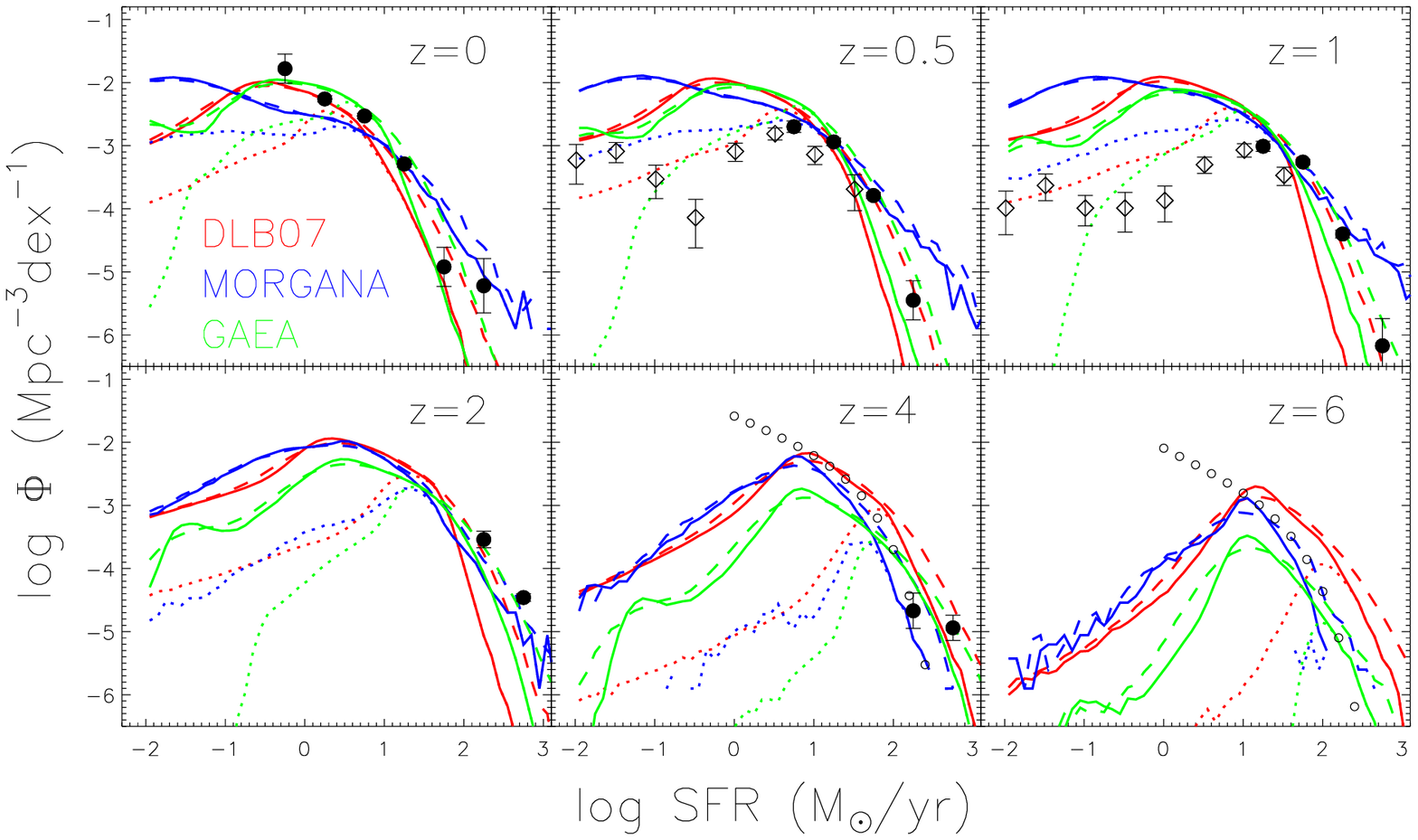}}\\%
\caption{ SFR functions at different redshifts for the three models considered in this study. Observational estimates are shown by symbols, and are taken from \citet[][black filled dots]{gruppioni2015}, \citet[][diamonds]{fontanot2012}, and \citet{mashian2016} (open circles, in this case, we are showing their fitting results). When necessary, data have been converted to a Chabrier IMF for consistency with model assumptions. Red, blue and green solid lines show results from DLB07, MORGANA and GAEA, respectively.  Solid lines are obtained considering all galaxies more massive than $10^9M_{\odot}$; dashed lines correspond to the same model predictions but convolved with an uncertainty of $\sim 0.3$~dex for the SFR; dotted lines show model results obtained considering all galaxies more massive than $10^{10}M_{\odot}$.} 
\label{fig:SFRF}
\ec
\end{figure*}

In this section, we analyse the SFR distributions of model galaxies as a function of cosmic times, and compare model predictions with recent observational estimates.  We start from the evolution of the galaxy SFR function and cosmic SFR density, that have been considered in earlier studies based on semi-analytic models \citep{fontanot2012, mitchell2014, gruppioni2015,hirschmann2016}. We then present the SFR - stellar mass relations, and the fraction of starburst galaxies as a function of redshift. In the following analysis, we include only model galaxies more massive than $10^9M_{\odot}$, which is above the resolution limit of all three models considered in this study. 

As discussed in Section 2, in the models, starbursts are driven by galaxy-galaxy mergers (and also by disk instabilities in MORGANA). For the following analysis, we will highlight the contribution of merger galaxies, defined as those that have experienced mergers during the time interval between two subsequent snapshots (this is of the order of 0.3 Gyr for z $<$ 3, and becomes less than 0.1 Gyr for z$>$5). Throughout the paper, we define major mergers as those characterized by a baryonic (stellar plus cold gas) mass ratio larger than 0.3. All other mergers with mass ratio smaller than 0.3 are defined as minor.

\subsection{SFR function and SFR density evolution}
Fig.~\ref{fig:SFRF} shows the SFR function (SFRF) at six different redshifts. Symbols show different observational estimates. Black solid dots with error bars are based on IR+UV flux-limited data from the PEP GOODS-S and COSMOS fields \citep{gruppioni2015}. Diamonds with error bars show the estimates by \citet{fontanot2012}, based on 24-$\mu$m data and corresponding to a mass-limited sample with $M_{stars}>10^{10}M_{\odot}$ from the GOODS--MUSIC survey. At z=4 and z=6, we show dust-corrected SFRFs by \citet{mashian2016}, based on the UV luminosity functions from the HST Legacy Fields \citep{bouwens2015}. Open circles show their Schechter fits to the data. Solid lines in Fig.~\ref{fig:SFRF} show model predictions corresponding to galaxies more massive than $10^9M_{\odot}$, dashed lines show these same predictions but assuming an uncertainty of $\sim 0.3$~dex for the SFR, and dotted lines show model results obtained considering only model galaxies more massive than $10^{10}M_{\odot}$. The latter can be compared directly with estimates by \citet{fontanot2012}, that are complete down to the same mass limit.  

In general, models predict a Schechter-like distribution for the SFRs, with a turn-over at low values that depend on the mass limit of the sample. In all three models considered in this study, the amplitudes of the SFRF at high SFR values peaks at about z=2. At $z < 1$, MORGANA predicts larger (lower) number densities of high (low) SFR galaxies than DLB07 and GAEA. Predictions from DLB07 and GAEA are close to each other at $z<1$,  while at higher redshift the overall normalization of the SFRF predicted by GAEA is lower than for the other two models considered \citep[see also][]{hirschmann2016}.

\begin{figure}
\bc
\hspace{-0.5cm}
\resizebox{8.cm}{!}{\includegraphics{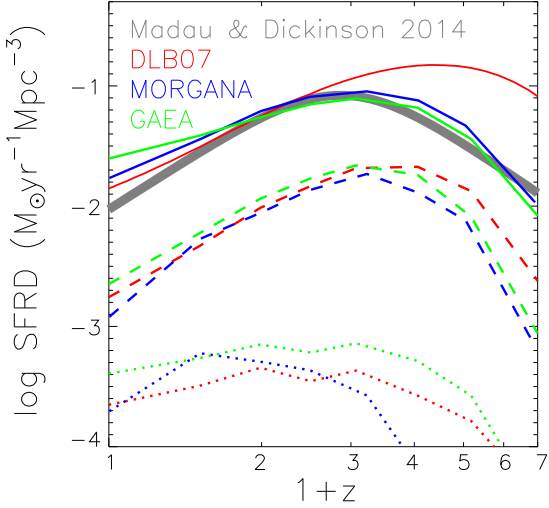}}\\%
\caption{SFR density evolution of all model galaxies (solid lines), galaxies that have experienced mergers (both minor and major) during the last snapshot interval (dashed lines), and galaxies that have experienced only major mergers during the last snapshot interval (dotted lines). The SFR densities plotted include stars formed during both starbursts and quiescent star formation.  As a reference, the thick gray line overplotted corresponds to the best-fit double-power-law function for the compilation of infrared to far-UV data by \citet{madau2014}, converted to a Chabrier IMF.} 
\label{fig:SFRD}
\ec
\end{figure}

In Fig.~\ref{fig:SFRD} we show the cosmic SFR density evolution predicted by all three models considered in this study. Solid lines correspond to predictions obtained considering all model galaxies more massive than $10^9M_{\odot}$; dashed lines are for galaxies that experienced mergers (both major and minor) during the last snapshot interval; dotted lines are for galaxies that experienced major mergers. The SFR values considered include both quiescent and merger/instability driven star formation, so dashed and dotted lines represent upper limits of the contribution of the starburst mode to the cosmic SFR density.

At $z=0$, GAEA predicts the highest SFR density, and DLB07 the lowest. At redshifts larger than $\sim 2$, DLB07 predicts much higher SFR density than the other two models. Although GAEA generally predicts larger gas masses than DLB07 at high redshift, both the SFRF and the SFR density at $z>2$ are significantly lower, due to the stronger feedback at earlier cosmic epochs \citep{hirschmann2016}. The difference between the dashed lines is smaller than that found between the solid lines, but the relative difference between the three models considered is the same as that discussed for all galaxies. In all three models considered, the dashed lines are offset about 0.6-1 dex below the solid lines, while the contribution of galaxies with major mergers to the total SFR density is only about 1 per cent or lower. Therefore,  the contribution from merger-induced starburst mode to the total SFR density is generally minor, at all cosmic epochs.  The same stands for the model of \citet{bower2006} as shown by \citet{lagos2011}. However, different results have been found for other semi-analytic models, where the contribution from starburst mode can become even larger than that of the quiescent mode at redshift larger than $\sim 3$ \citep{baugh2005, gp2014}, or $\sim 5.5$ \citep{benson2010}.  \citet{lagos2011} shows that in the models where longer SF time-scales or lower SF efficiencies are assumed in the quiescent mode, disk instabilities can make the contribution from starburst significant because of the larger amount of gas typically involved.

\subsection{SFR - stellar mass relation}

\begin{figure*}
\bc
\hspace{-1.4cm}
\resizebox{16cm}{!}{\includegraphics{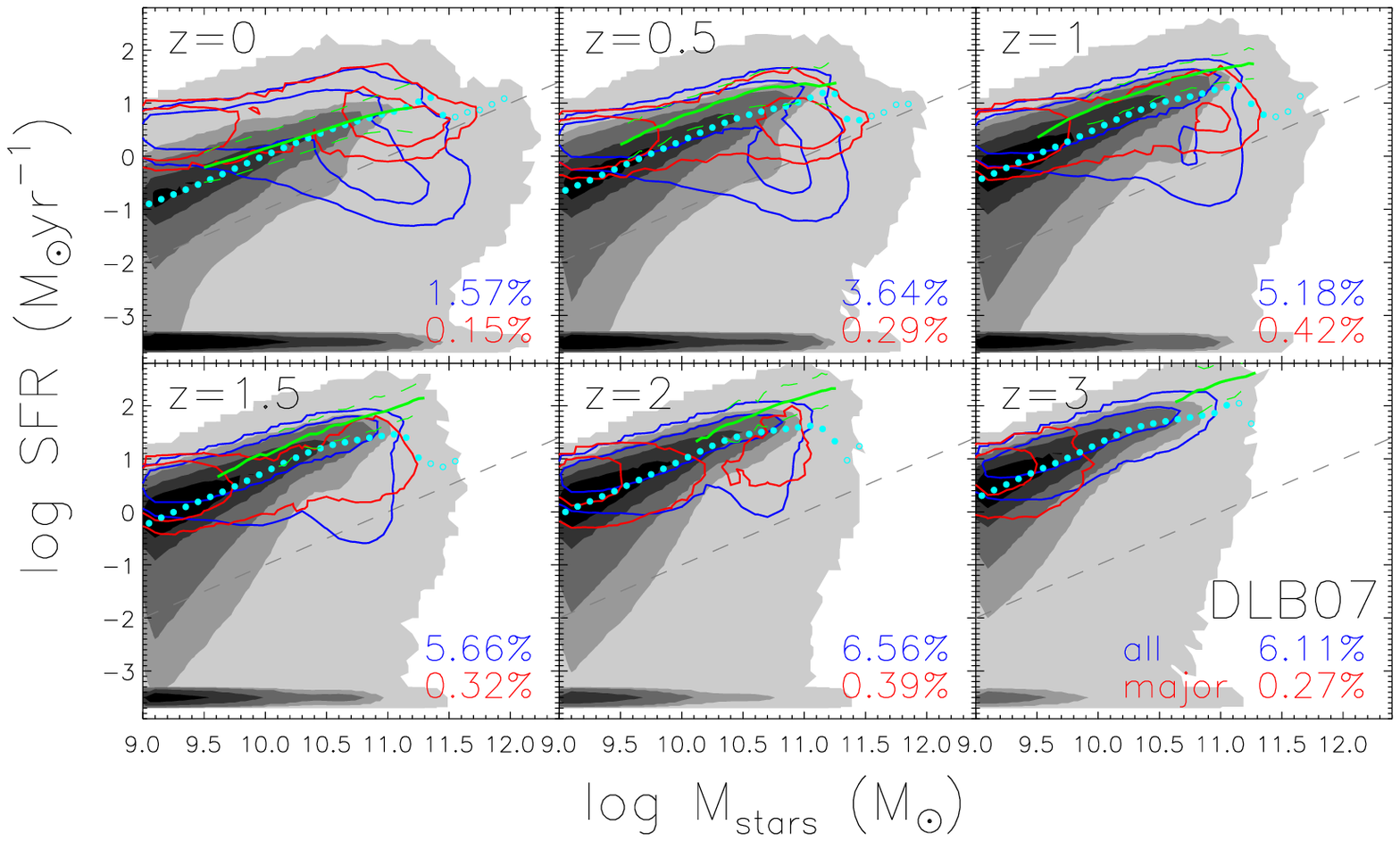}}\\%
\caption{SFR-stellar mass relation predicted by the DLB07 model, at six different redshifts. Green lines show the observational estimates by \citet{schreiber2015}, with solid green lines showing the average stacked SFR, and dashed green lines showing the 1$\sigma$ dispersion. In each panel, dark shaded regions show the distribution of model galaxies, with contours including 38, 68, 87, 95 and 100 per cent of all model galaxies going from black to light gray, respectively. Blue contours show the distribution of galaxies that experienced mergers (both minor and major) at each redshift, with the two levels corresponding to regions including 68 and 95 per cent of the entire sample. Finally, red contours show the distribution of model galaxies that experienced major mergers. Cyan circles show the SFR main sequence ($SFR_{\rm MS}$) predicted by the model, defined as the median SFR of star forming galaxies (those with $\log$(sSFR) larger than -11, i.e., above the gray dashed line in each panel). Filled cyan circles are used for stellar mass bins where the number of galaxies included is larger than 200, while open cyan circles represent bins with less than 200 galaxies. The  blue/red numbers listed in each panel give the fractions of merger/major merger galaxies at each redshift.}
\label{fig:SFRMstars}
\ec
\end{figure*}

\begin{figure*}
\bc
\hspace{-1.4cm}
\resizebox{16cm}{!}{\includegraphics{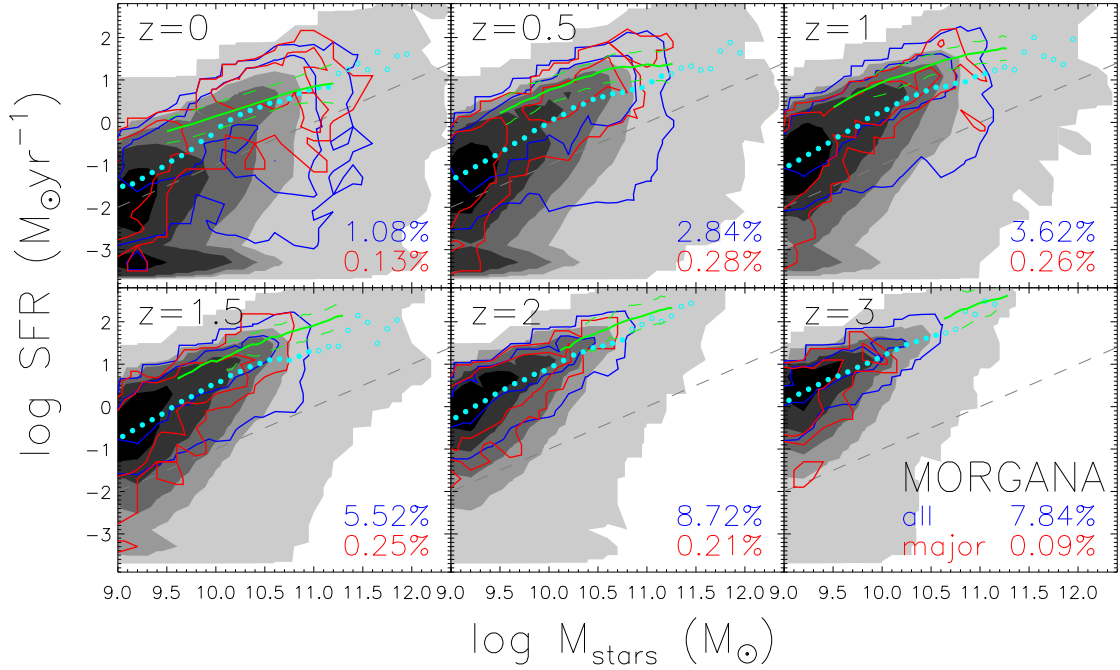}}\\%
\caption{ As for Fig. 3, but for the MORGANA model. } 
\label{fig:SFRMstars_Morgana}
\ec
\end{figure*}

\begin{figure*}
\bc
\hspace{-1.4cm}
\resizebox{16cm}{!}{\includegraphics{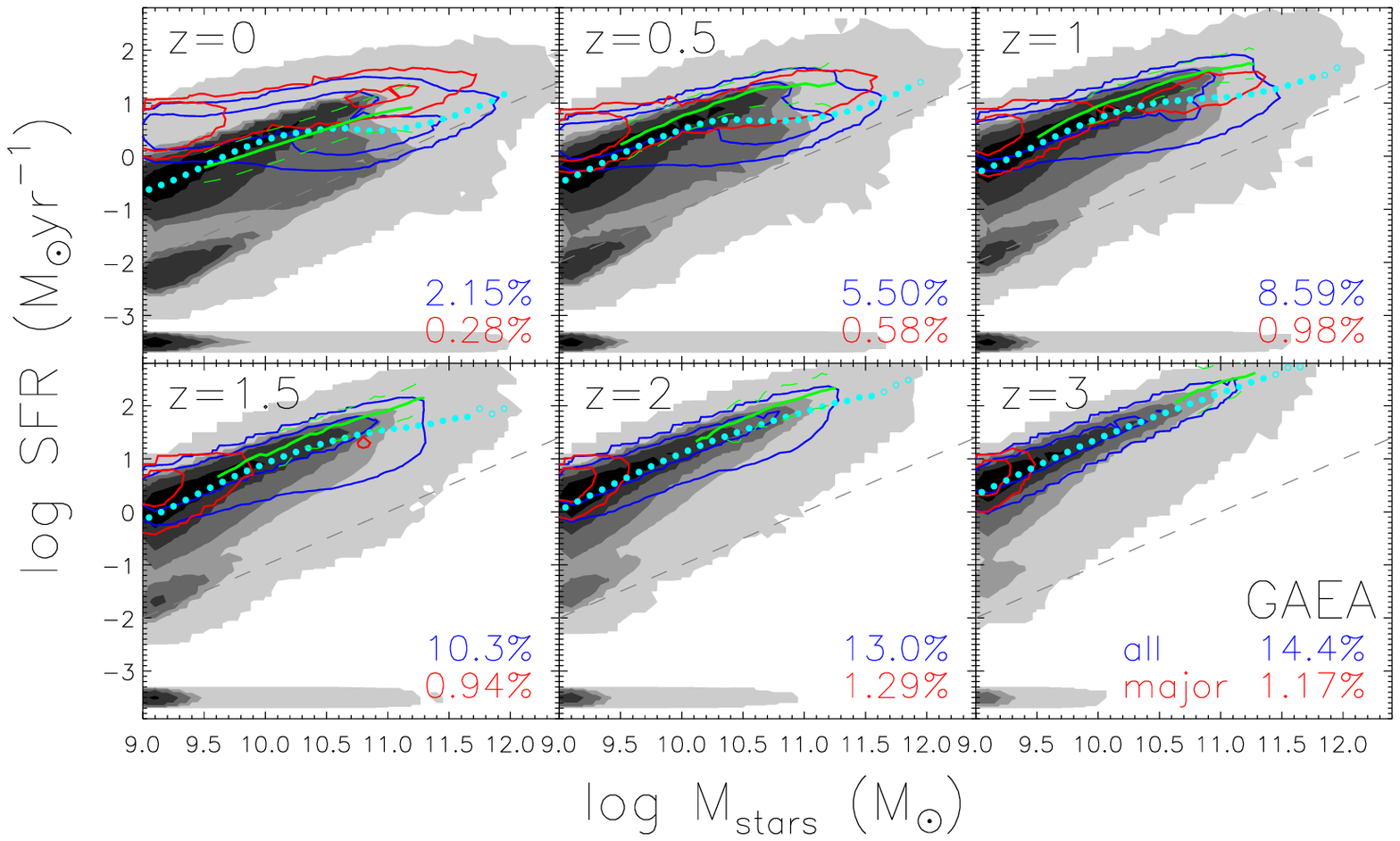}}\\%
\caption{ As for Fig. 3, but for the GAEA model. } 
\label{fig:SFRMstars_GAEA}
\ec
\end{figure*}

Figs.~\ref{fig:SFRMstars},~\ref{fig:SFRMstars_Morgana} and~\ref{fig:SFRMstars_GAEA} show the SFR-M$_{\rm stars}$ relations predicted by the DLB07, MORGANA, and GAEA model, respectively. For visualization purposes, all model galaxies with $\log$SFR$(M_{\odot}yr^{-1}) <-3.5$ have been assigned to have $\log$SFR=-3.5. Black to light gray regions include 38, 68, 87, 95 and 100 per cent of all model galaxies at each redshift considered. The solid green line in each panel shows the average SFR of star forming galaxies (selected according to their location in the UVJ colour-colour diagram) as estimated by \citet{schreiber2015}. The dashed lines show the 1$\sigma$ dispersion of the observational estimates. For the models, we define a SF main sequence (SFR$_{MS}$) as the median SFR of all galaxies with $\log($sSFR$) > -11$  (this is shown as a dashed grey line in each panel of Figs.~3-5). The resulting model main sequence is shown by cyan symbols in the figures. Predictions from all the three models considered in this study are in relatively good agreement with observational estimates at $z=0$, especially for galaxies with stellar mass larger than $\sim 10^{10}M_{\odot}$. At higher redshifts, the slope of the predicted MS is comparable to that observed, but all models tend to under-estimate the average level of SF observed, at all stellar masses.  
All models also predict, at all redshift, a significant population of passive galaxies with extremely low SFRs. In MORGANA, the star forming sequence has a significantly larger scatter and a steeper slope (for $z<2$) than the DLB07 model. In GAEA, in addition to the population of  passive galaxies with SFR $\le -3.5$, there is a secondary peak around  log SFR $\sim -2$, at all redshifts. The SF main sequence predicted by this model has a higher amplitude than the other two models considered, and is closer to the observational estimates. 

The blue(red) lines in the figures show the 68 and 95 percentile contours of all galaxies that have experienced mergers (major mergers) in the time interval between the redshift corresponding to each panel and the immediately previous snapshot. Low mass galaxies that have experienced mergers in the DLB07 model have generally larger SFR than the overall star forming population, and tend to reside above the model main sequence.  More massive galaxies that experienced mergers, however, cover a wider range of SFRs at  $z \le 2$, and most of the merger galaxies more massive than $\sim 10^{11}M_{\odot}$ have SFR lower than the main sequence.  These are likely gas poor (dry) mergers for which the lack of significant gas amounts does not lead to a significant starburst.  At $z \le 2$, major merger galaxies have similar distribution as all merger galaxies, but exhibit higher SFRs for massive galaxies. At $z=3$, the SFR-M$_{\rm stars}$ relations for the merger galaxies are similar as that of all galaxies, and major mergers are mostly confined to the least massive galaxies. In MORGANA, galaxies that experienced mergers have similar distributions to the overall star forming galaxy population, with no obvious excess of SFR. Their distribution also extends down to the passive population at the low mass end. In GAEA, most of the major merger galaxies reside above the model main sequence, at all redshifts and galaxy masses considered. The most massive galaxies that experienced mergers can still be found in an active phase, with SFRs larger than those measured for the average star forming galaxy population. This is a consequence of the fact that more gas is available in GAEA compared to DLB07 and MORGANA.

The actual fractions of merger and major merger galaxies with respect to the overall population at each redshift are indicated by the numbers given in the bottom right corner of each panel in Figs.~3-5. In the DLB07 model, the fraction of galaxies that experienced a merger event (both minor and major) increases from  $\sim 1.6$ per cent  at $z=0$ to about 6 per cent at $z > 2$. The fraction of galaxies that experienced major merger is generally quite low, always less than $\sim 0.5$ per cent.  In MORGANA, there are systematically less merger galaxies at $z \le 1$, and less major merger galaxies at all redshifts than in DLB07. In GAEA, the fractions of merger galaxies (more than 10 per cent at $ z>1$) are around 1.5-2 times, and the fractions of major merger galaxies (more than 1 per cent at $z>1$) are around 2-10 times those in DLB07 and MORGANA, at all redshifts.  

DLB07 and GAEA are based on the same dark matter halo merger trees and treat the dynamical evolution of satellite galaxy mergers in the same way.  Therefore, the difference in the fraction of merger galaxies found between these two models is due to the different amplitude of the stellar mass functions at the low mass end \citep{hirschmann2016}. 

\subsection{Fraction of starburst galaxies}

\begin{figure*}
\bc
\hspace{-0.8cm}
\resizebox{16cm}{!}{\includegraphics{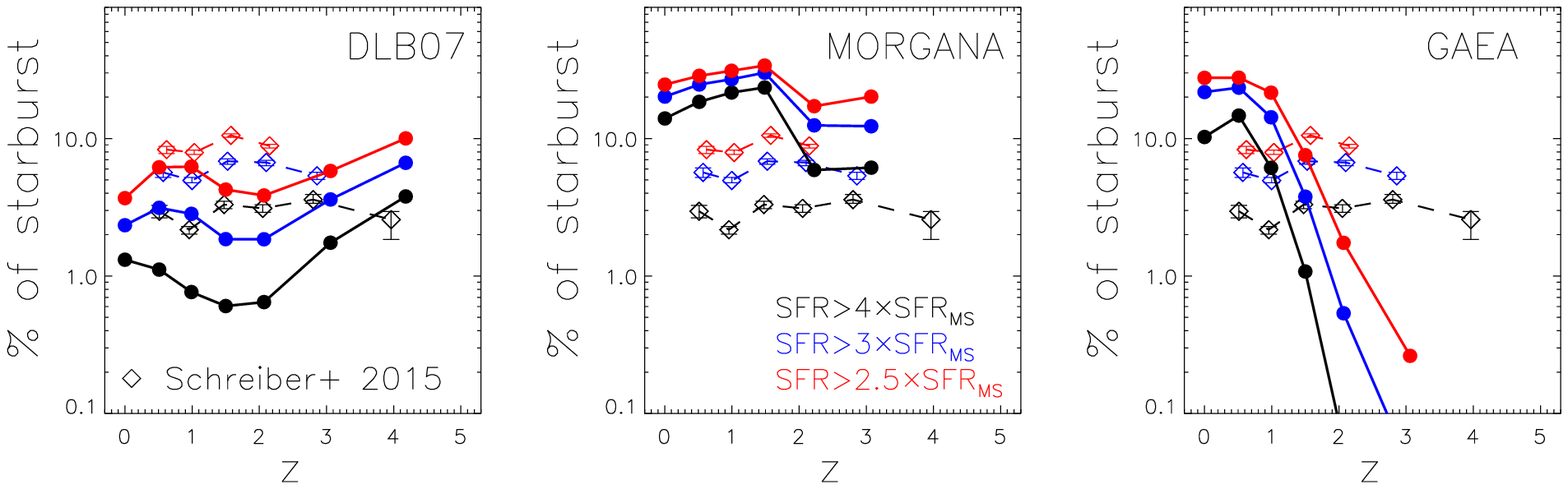}}\\%
\caption{ Fraction of starburst galaxies as a function of redshift in the three models considered in this study (from left to right: DLB07, MORGANA, and GAEA). Only galaxies more massive than   $5\times10^{10}M_{\odot}$ have been considered. Black, blue and red solid lines correspond to different thresholds above the model main sequence, as indicated in the legend. Diamonds with error bars connected by dashed lines show the observational estimates by  \citet[][Fig. 18]{schreiber2015}, corresponding to an approximately complete sample of galaxies more massive than $5\times10^{10}M_{\odot}$ at the redshifts investigated.} 
\label{fig:SBfraction}
\ec
\end{figure*}

As mentioned in Section 1, galaxies that lie above the observed main sequence are typically considered to be triggered by mergers and/or disk instabilities and classified as starburst galaxies. In Fig.~\ref{fig:SBfraction}, we compare the estimated starburst fraction with predictions from the three different models considered in this study, as a function of redshift. Diamonds connected by dashed lines show the observational estimates by  \citet{schreiber2015}, corresponding to an almost complete sample  (with completeness more than 90 per cent) of galaxies more massive than $5\times10^{10}M_{\odot}$  at the redshifts investigated. Starburst galaxies are defined as those with SFR more than X times the observed main sequence, where X = 2.5, 3, and 4 respectively. The estimated starburst fraction does not vary significantly as a function of redshift, but of course increases (albeit not significantly) when considering lower offsets from the main sequence.  

In Fig.~\ref{fig:SBfraction}, model predictions are shown by filled circles connected by solid lines and have been computed considering only galaxies more massive than $5\times10^{10}M_{\odot}$ (for consistency with observations), and with $\log$(sSFR) $> -11$.  Following the observational selections, we have classified as model starburst galaxies those with SFR larger than X times the main sequence defined by the star forming galaxies  of each model (cyan circles in the SFR-M$_{\rm stars}$ figures). At $z<1$, the DLB07 model predicts lower starburst fraction, while MORGANA and GAEA  higher starburst fraction than observed. As for the evolution as a function of redshift,  this is weak for DLB07 and MORGANA, while the starburst fraction drops dramatically at high redshifts for GAEA. This is possibly due to the strong outflows that suppress star formation at high redshifts. 

\section{Are starburst galaxies merger galaxies? }
\label{sec:starburst}

From the previous section, it is clear that for all the three models considered, galaxies that have experienced recent mergers can have relative low SFRs, and would not be classified as starburst galaxies using the selection typically adopted in observational data analysis. In other words, not all `starbust mode' galaxies, i.e. those that experienced merger driven bursts of star formation (or instability driven bursts in MORGANA) would be classified as starburst. Turning the question around, we can ask what is the fraction of galaxies that reside significantly above the MS (that would be classified as starbursts observationally) that experienced a recent merger or an instability-driven burst of star formation.  We address this question in the following. 

\subsection{Starburstiness distribution}

\begin{figure*}
\bc
\hspace{-1.4cm}
\resizebox{16.cm}{!}{\includegraphics{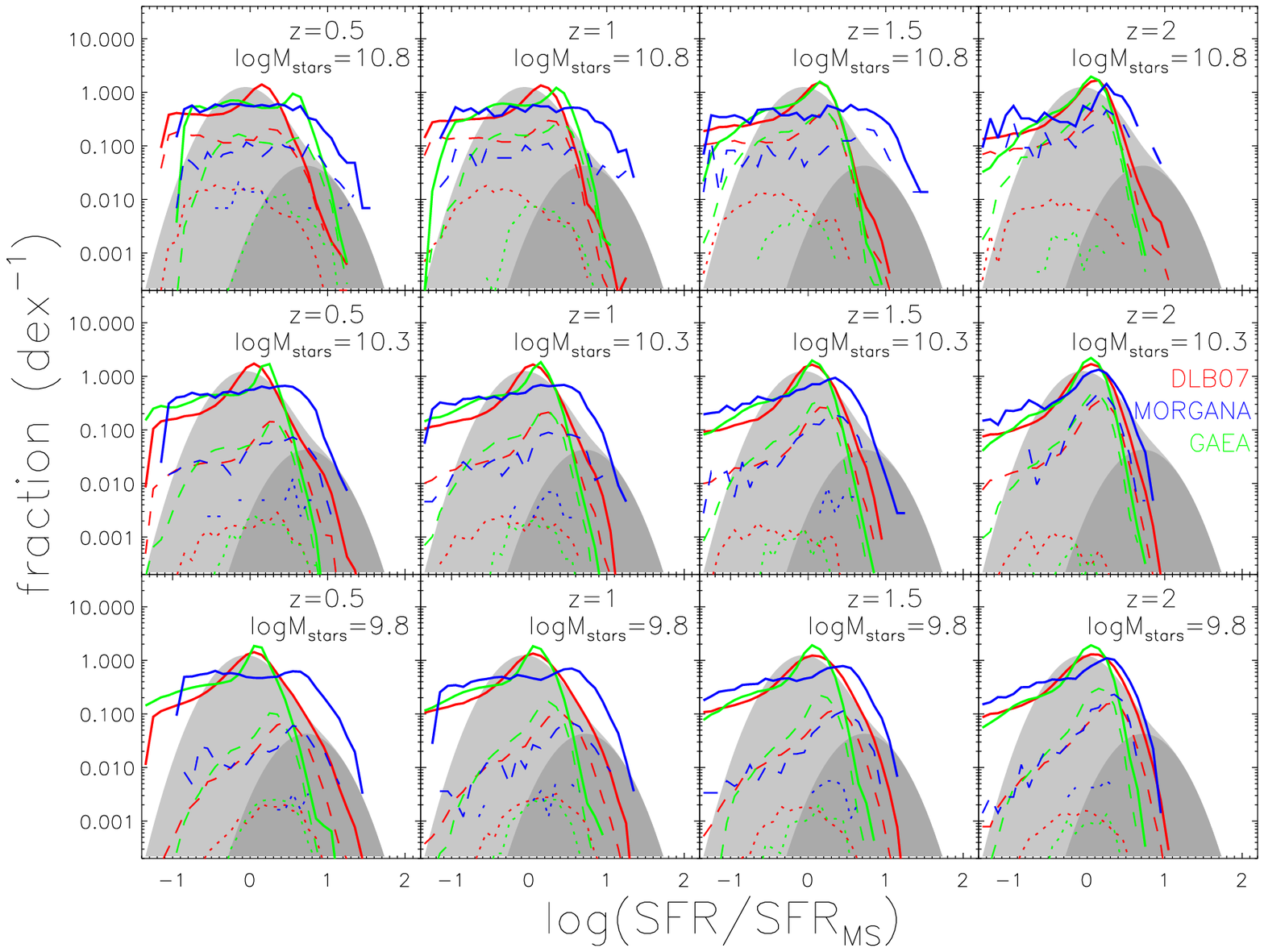}}\\%
\caption{ Starburstiness(SFR/SFR$_{\rm MS}$) distribution of star forming galaxies for different redshift and stellar mass bins, as labeled in each panel. Red, blue, and green lines show results from DLB07, MORGANA and GAEA, respectively. Solid lines show model predictions for all star forming galaxies; dashed lines are for galaxies that experienced recent mergers (including both minor and major mergers); and dotted lines are for galaxies that experienced recent major mergers. The gray regions, identical in all panels,  correspond to the best fits obtained for the observed starburstiness distribution as given by \citet{schreiber2015}. The light gray region corresponds to the entire observational star forming sample, while the dark gray region represents the `starburst bump'.} 
\label{fig:SBness}
\ec
\end{figure*}

For a given galaxy, a `starburstiness' can be defined as the ratio between galaxy SFR and the main sequence SFR value (SFR$_{\rm MS}$) corresponding to the stellar mass and redshift of the galaxy \citep{elbaz2011}. This is then a measure of the offset above the main sequence. \citet{schreiber2015} found that the starburstiness distributions of galaxies exhibits a universal shape that does not depend significantly on redshift and galaxy mass, and is well described by a Gaussion with a secondary `starburst bump'. The starburstiness distributions can be fit using a double log-normal distribution \citep[][Fig.~17, Fig.~19 and Equation 10]{schreiber2015}.

In Fig.~\ref{fig:SBness}, we show the starburstiness distribution predicted for star forming galaxies in the three models considered in this study. Different panels correspond to different redshifts and galaxy stellar mass bins, as indicated in the legend. Gray shaded regions show the best fit to the observational distributions; solid lines show the distributions obtained considering all model star forming galaxies; dashed and dotted lines are for galaxies that recently experienced a merger (both minor or major) and a major merger, respectively. 

For the DLB07 model (red solid lines), the distribution of offsets from the MS exhibits somewhat a `bump' for low redshifts and low stellar masses. GAEA (green solid lines) has a distribution that is very similar to that predicted by the DLB07 model around the main sequence peak but the overall distribution is narrower with less galaxies at large offsets from the MS. For MORGANA, the distributions are flatter and wider than found for DLB07 and GAEA in most of the panels, consistently with the trends seen considering the SFR-M$_{\rm stars}$ distributions. Comparing the solid and dashed lines at high values of starburstiness (large offsets from the MS) quantifies how many high SFR galaxies are triggered by recent mergers. The figure shows clearly that merger galaxies contribute to only a part of the galaxies with largest deviations from the main sequence, for all three models considered in this study. In other words, many starburst galaxies are `normal' galaxies that did not experience a merger in their recent past. The contributions of merger galaxies to galaxies with the highest values of  SFR are smallest in MORGANA, and higher in DLB07 and GAEA. The fractions of merger galaxies in starburst galaxies are shown in the next subsection.

\subsection{Fraction of starburst galaxies associated with merger events}

\begin{figure*}
\bc
\hspace{-1.4cm}
\resizebox{16cm}{!}{\includegraphics{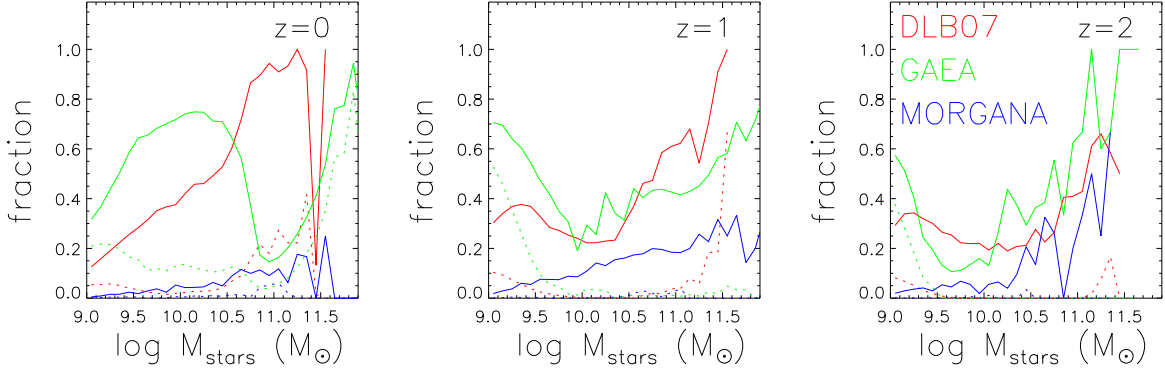}}\\%
\caption{For starburst galaxies that have SFR values larger than 4 times the median value of star forming galaxies, the fraction of merger (solid lines) and major merger (dotted lines) galaxies at given galaxy mass. As indicated in the legend, different panels correspond to different redshifts while different colours are used for the three models considered in this study.} 
\label{fig:checkSB}
\ec
\end{figure*}

Fig.~\ref{fig:checkSB} quantifies the fraction of starburst galaxies that experienced a recent merger as a function of galaxy stellar mass, for the three models considered in this study. Here we have defined as starburst galaxies  those that have SFR larger than 4 times that corresponding to the model main sequence. Solid lines show fractions for all mergers (minor and major), while dotted lines are obtained when considering only galaxies that experienced a recent major merger event. The fractions obtained are always well below 1, i.e. in all models considered only a minority of the galaxies offset above the main sequence have elevated values of SF because of a recent merger event. 

For the DLB07 model, the fraction of merger galaxies contributing to the starburst population increases with stellar mass. MORGANA exhibits a similar dependence on galaxy stellar mass, but predicts overall smaller fractions.
We recall that MORGANA has an extra channel for starburst, i.e. disk instabilities that are not triggered by mergers.
In GAEA, the fractions of merger galaxies are close to those predicted by the DLB07 model at  z=1 and z=2. At z=0, for galaxies less massive than $\sim 10^{10.5} M_{\odot}$, the merger fraction is higher in GAEA than in DLB07, while it drops dramatically for galaxies with mass around $\sim 10^{11} M_{\odot}$.  This is very much sensitive to the way we have defined the model main sequence.  Fig.~5 shows that most galaxies of this stellar mass lie around the dashed gray line at  z=0 and z=0.5,  and have $\log$(sSFR) $\sim -11$. As a result, the model main sequence defined as the median SFR of active galaxies shows a `dip' around $10^{11} M_{\odot}$. The peak of the starburstiness distribution therefore shifts towards higher values compared to DLB07 (e.g., left top panel in Fig. 7), and the starburst galaxy population includes many more quiescent galaxies than for DLB07.  In all the three models considered, major mergers (dotted lines) account for a very small fraction of  the starburst galaxies, with somewhat larger fractions (but still below $\sim 1$ per cent) for the lowest and/or most massive galaxies in DLB07 and GAEA.

\section{Why not all merger galaxies are starburst? And why some quiescent galaxies are?}
\label{sec:check}

The analysis illustrated in the previous sections has clarified that not all model galaxies that experienced a recent merger or even a recent major merger episode have SFRs above the main sequence. On the other hand, there are a number of starburst galaxies (with SFR values above the main sequence) that are not associated with a recent merger episode. In this section, we analyse in more detail the origin of these findings. 

\subsection{Relation between cold gas mass and starburstiness}

\begin{figure*}
\bc
\hspace{-0.8cm}
\resizebox{16cm}{!}{\includegraphics{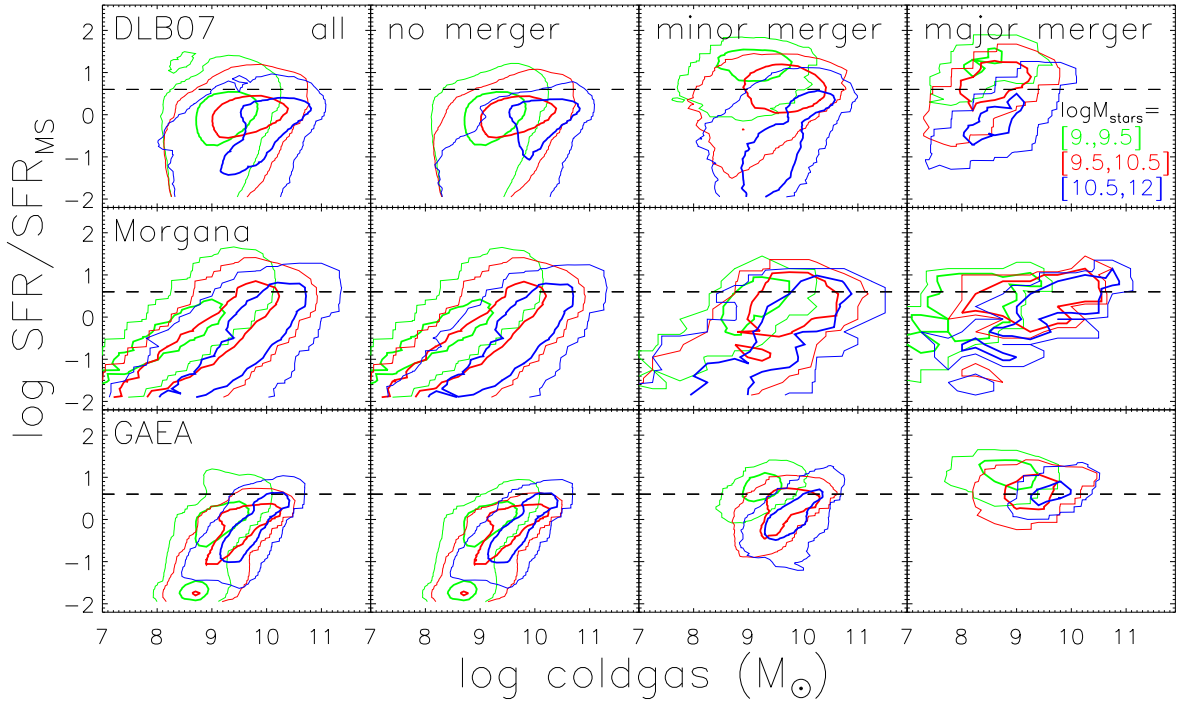}}\\%
\caption{Relation between starburstiness (offset from the main sequence) and cold gas mass for galaxies in the three models used in this study (different rows), at z=0. Different colours correspond to different stellar mass bins, as indicated in the legend. Panels from left to right show all galaxies, galaxies with no recent mergers, galaxies that experienced a recent minor merger, and galaxies that experienced a recent major merger. The two contours plotted in each panel and for each stellar mass bin correspond to contours containing 68 (thick contours) and 99 (thin contours) per cent of the galaxies in each of the samples considered. Dashed horizontal lines indicate the threshold for considering a galaxy as a starburst (SFR larger than four times that of the main sequence).} 
\label{fig:checkmerger}
\ec
\end{figure*}

To understand what drives the SFRs of galaxies that experienced/did not experience recent mergers, we first consider the relation between the galaxy cold gas mass and starburstiness in the three models adopted.  Results are shown  in Fig.~\ref{fig:checkmerger}, for galaxies at z=0 in different stellar mass bins. At higher redshifts, results are qualitatively the same. Panels from left to right in Fig.~\ref{fig:checkmerger} show the distribution for (i) all galaxies, (ii) galaxies that did not experience a merger in the last snapshot interval, (iii) galaxies that experienced a recent minor merger (and no major merger), and (iv) galaxies that experienced a recent major merger. Different rows correspond to different models, while different colours correspond to different galaxy stellar mass bins. Fig.~\ref{fig:checkmerger} shows again that even for major merger galaxies, not all of them are starburst galaxies with high starburstiness. At z=0, the fraction of major merger galaxies that are starburst are 58, 25 and 82 per cent in DLB07, MORGANA and GAEA respectively.

For all three models considered, there is an overall trend for galaxies with larger cold gas mass to have higher starburstiness, i.e. to deviate more from the main sequence.  
For galaxies that had no recent merger, the distribution becomes almost vertical towards low cold gas masses for DLB07 and GAEA. This is due to the assumption of a critical threshold for star formation in these models.  In MORGANA, no critical cold gas mass limit is set, and therefore the correlation between cold gas mass and starburstiness extends to lower cold gas masses.

In DLB07 and GAEA, massive galaxies  with only minor mergers do not exhibit a much larger starburstiness than galaxies without recent mergers; a more significant difference is found for low mass galaxies.  Galaxies that experienced a recent major merger in DLB07 and GAEA have larger starburstiness than the ones without any recent merger. In MORGANA, galaxies with mergers have similar distribution as galaxies with no mergers, except when considering galaxies in the lowest mass bin that experienced a recent major merger. 

To summarize, in all models considered in this study, for galaxies that did not experience a recent merger, the level of SFR is driven by the amount of cold gas available. The scatter is quite large, which reflects a scatter both in cold gas mass and in physical radii, and some galaxies can be offset significantly above the main sequence. While in all models considered mergers trigger a burst of star formation, the amount of gas involved can be insufficient to lead to a significant enhancement of the SFR. Therefore, a typical selection employed in observational studies would result in a sample of starburst galaxies that include a number of galaxies that are forming stars in a `quiescent mode' and exclude a number of systems that experienced a recent merger in their past. 

\subsection{Starburst galaxies with no recent merger}

In the previous subsection, we have demonstrated that for a fraction of the galaxies that deviate significantly from the main sequence, the star formation is not triggered by a recent merger episode. In this subsection, we analyse how these galaxies are driven to elevated rates of star formation, and what is the origin of their cold gas reservoir. 

\begin{figure*}
\bc
\hspace{-0.8cm}
\resizebox{16cm}{!}{\includegraphics{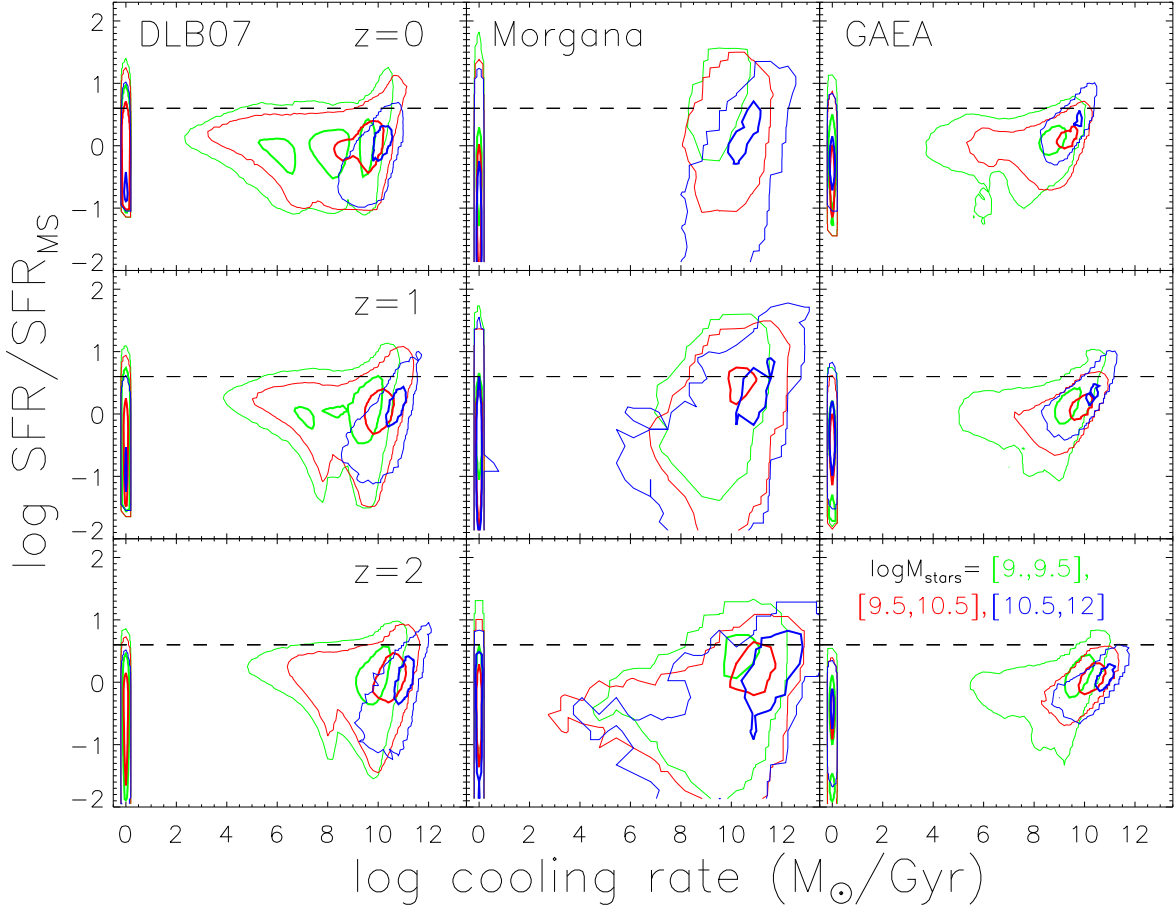}}\\%
\caption{  For the three models and at three redshifts, the relations between the rate of gas cooling during the last snapshot interval (cooling gas mass divided by the time interval between the two snapshots, in unit of  $M_{\odot}$/Gyr) and starburstiness ($SFR/SFR_{\rm MS}$), for galaxies that did not experience a merger in their recent past. Different colours correspond to different galaxy stellar mass bins, and contours enclose 68 and 99 per cent of the galaxies in each sub-sample. The horizontal dashed line shows the starburstiness level assumed to classify galaxies as starburst (four times above the main sequence).} 
\label{fig:nomerger_cooling}
\ec
\end{figure*}

In Fig.~\ref{fig:nomerger_cooling}, we present the relation between the starburstiness and the rate of gas cooling (cooling gas mass during the last snapshot interval divided by the time interval between the two snapshots, in unit of $M_{\odot}$/Gyr ), for galaxies with no recent merger and at three redshifts. In the models considered in this study, cold gas mass  can increase either via mergers/accretion with/of gas rich satellites (these are excluded by construction in this case) or via gas cooling.  In Fig.~\ref{fig:nomerger_cooling}, galaxies above the horizontal dashed lines would be classified as starburst. The figure shows that in all three models, many of these starburst galaxies have a lot of gas that originates from cooling, but there is also a population of starburst galaxies with little or no cooling gas. The fraction of non-merger starburst galaxies with cooling gas less than $10 M_{\odot}$/Gyr decreases with increasing redshift. Specifically, it amounts to about 62 (41, 42) per cent in DLB07 (MORGANA, GAEA) at z=0, and it decreases to $\sim$ 7 (21, 0.3) per cent at z=2. 



As described in Section 2.1, a critical cold gas mass threshold is assumed in DLB07 and GAEA, with this threshold depending on the galaxy disk radius. In MORGANA, no threshold is assumed for star formation to take place, but the SFR is proportional to the cold gas surface density, which also depends on disk radius. We have checked that in all the three models we study, non-merger starburst galaxies tend to be dominated by galaxies with small disk radius.

\begin{figure*}
\bc
\hspace{-1.4cm}
\resizebox{16cm}{!}{\includegraphics{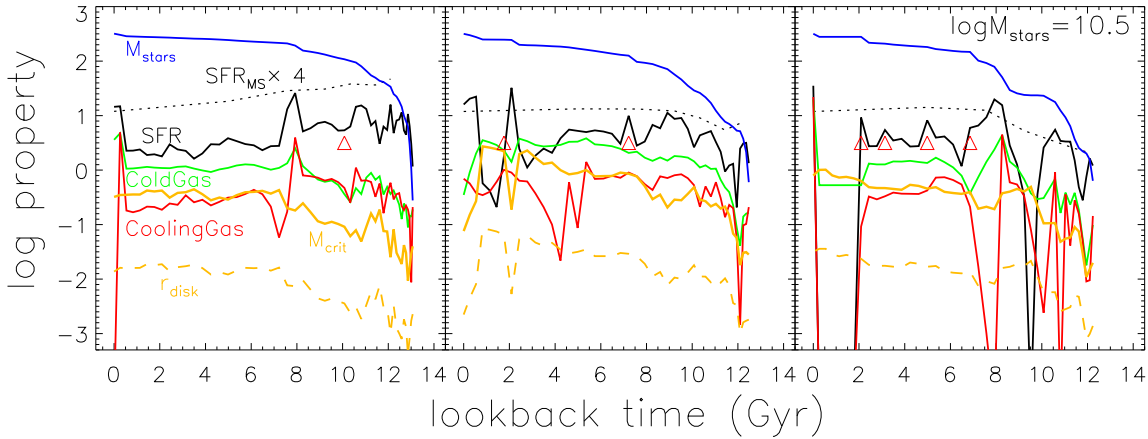}}\\%
\caption{Formation histories of three representative non-merger starburst galaxies with stellar mass of $10^{10.5}M_{\odot}$, from the DLB07 model. Lines of different colors indicate the evolution of different galaxy properties as a function of look back time: blue line - stellar mass (in units of $10^8M_{\odot}$); black solid line - SFR ($M_{\odot}$/yr); green line - cold gas mass ($10^{10}M_{\odot}$); red line - cooling gas mass during the last snapshot interval ($10^{10}M_{\odot}$); orange dashed line - disk radius (Mpc). For each galaxy, the orange solid line corresponds to the critical cold gas mass threshold ($10^{10}M_{\odot}$) at the corresponding lookback time, and the black dotted line shows a value corresponding to 4 times the SFR main sequence. Red triangles mark the times corresponding to mergers (all minor mergers in this case).} 
\label{fig:SB_his}
\ec
\end{figure*}

What is the reason why a significant fraction of non-merger starburst galaxies have absolutely no gas cooling in the last snapshot interval?
What is the relative importance of cooling gas, cold gas, and disk radius in determining the SFR of a starburst galaxy?  To address these questions, we analyse  the history of non-merger starburst galaxies at z=0 in the DLB07 model. A few examples are shown in Fig.~\ref{fig:SB_his}, for galaxies of mass $10^{10.5}M_{\odot}$ at z=0. Galaxy properties at different look back times, including SFR, stellar mass, cold gas mass, cooling gas mass, and disk radius are plotted using different color lines. For each galaxy, the orange solid line shows the critical cold gas mass threshold $M_{\rm crit}$ at the corresponding lookback time, and the black dotted line shows a value corresponding to 4 times the SFR of main sequence galaxies in the model (at the stellar mass corresponding to the redshift considered). 

Fig.~\ref{fig:SB_his} shows that the SFR (black solid lines) is determined by both cold gas mass (green lines)  and $M_{\rm crit}$ (orange lines), as expected. When the cold gas mass is smaller than $M_{\rm crit}$ (see the right panel at look back time of about 1Gyr), there is no SF in the galaxy. When sufficient cold gas mass is available, the SFR is higher when there is more cold gas, or when there is an obvious decrease of $M_{\rm crit}$. Cold gas mass can increase significantly when a lot of gas is cooling (red lines), and $M_{\rm crit}$ is closely related to the disk radius of the galaxy (orange dashed lines). In the example shown in the right panel, the galaxy becomes a starburst at z=0 due to a large increase in cold gas mass that comes from an increase of the gas cooling. In the left panel, the increase of cooling gas happens a few snapshots before z=0, and the galaxy remains in a starburst phase down to present because of the large amount of cold gas available. In the middle panel, although the cold gas mass decreases towards z=0, there is an even stronger decrease of $M_{\rm crit}$ due to a decrease of the galaxy disk radius, which results into a burst of SF at present.  

To summarize, significant bursts of star formation can occur in non-merger galaxies when there is either high cold gas mass due to significant cooling, or a rapid decrease of the critical mass threshold as a consequence of a decrease of the gas disk radius. On the other hand, although all mergers (marked by red triangles in Fig.~\ref{fig:SB_his}) trigger a burst of SF in the models considered in our study, these might not be large enough (because of the low amount of gas involved) to offset significantly the galaxy from the main sequence. 

\section{Conclusions and discussions}
\label{sec:conclusion}
In this work,  we analyse the properties of star forming and starburst galaxies in three independently developed semi-analytic models of galaxy formation (DLB07, MORGANA and GAEA). All three models considered implement two modes of star formation: (i) a `quiescent' mode that takes place from cold gas associated with galaxy disks, and (ii) a `starburst' mode that is associated with galaxy mergers. In one of the models (MORGANA) `starburst'  events can also be triggered by disk instability episodes. 

In all the three models considered, we find that galaxies that experienced recent mergers (both minor and major mergers) contribute to only a very small fraction of the total cosmic star formation history. As for the relation between mergers and starburst, we find that not all galaxies with recent mergers are offset significantly above the main sequence (i.e. are starburst galaxies that have SFRs larger than 4 times the median value of star forming galaxies at given stellar mass).  Note that we do not set a minimum mass ratio to define minor mergers.  Therefore, it is not surprising that not all minor merger galaxies are starburst, because mergers with very small mass ratios are not expected to cause a starburst. Nevertheless, we find that even for major merger galaxies, not all of them are associated with a starburst. At z=0, the fraction of major merger galaxies that are starburst are 58, 25 and 82 per cent in DLB07, MORGANA and GAEA respectively.
Some merger galaxies can have relatively low rates of star formation because the merger is gas-poor. 

On the other hand, not all starburst galaxies experienced recent merger events. These non-merger but starburst galaxies actually contribute to a large fraction (that can reach $\sim$50 per cent or even more) of the starburst population at stellar masses less than $\sim 10^{10-10.5}M_{\odot}$. Their elevated rates of star formation can be generally ascribed to large cold gas masses (due to cooling and/or recent accretion events) and/or small disk radii, which leads to a higher cold gas surface density and therefore higher levels of star formation.

Whether the increase of cold gas and the decrease of disk radii -- that lead to starbursts in non-merger galaxies--are physical, and exist in the real Universe, in similar amount as in the model, remains unclear. 
Improved modeling of disk radii has been recently proposed in several semi-analytic models \citep[e.g.][]{xie2017}. In future work, we will study the influence of these different assumptions on the origin of the predicted starburst population.


\section*{Acknowledgements}

We thank Corentin Schreiber for providing us data from his paper in electronic format. LW acknowledges support from the NSFC grants program (No. 11573031), and the National Key Program for Science and Technology Research and Development (2017YFB0203300). 
GDL acknowledges financial support from the MERAC foundation.  MH acknowledges financial support from the European Research Council (ERC) via an Advanced Grant under grant agreement no. 321323–NEOGAL.


\bibliographystyle{mn2e}
\bibliography{SFR}

\label{lastpage}

\end{document}